# Tracing Execution of Software for Design Coverage

Raimondas Lencevicius    Edu Metz    Alexander Ran

Nokia Research Center, 5 Wayside Road, Burlington, MA 01803, USA

Raimondas.Lencevicius@nokia.com    Edu.Metz@nokia.com    Alexander.Ran@nokia.com

**ABSTRACT**

Test suites are designed to validate the operation of a system against requirements. One important aspect of a test suite design is to ensure that system operation logic is tested completely. A test suite should drive a system through all abstract states to exercise all possible cases of its operation. This is a difficult task. Code coverage tools support test suite designers by providing the information about which parts of source code are covered during system execution. Unfortunately, code coverage tools produce only source code coverage information. For a test engineer it is often hard to understand what the noncovered parts of the source code do and how they relate to requirements. We propose a generic approach that provides design coverage of the executed software simplifying the development of new test suites. We demonstrate our approach on common design abstractions such as statecharts, activity diagrams, message sequence charts and structure diagrams. We implement the design coverage using Third Eye tracing and trace analysis framework. Using design coverage, test suites could be created faster by focussing on untested design elements.

## 1    INTRODUCTION

Test suites are designed to validate the operation of a system against requirements. One important aspect of a test suite design is to ensure that system operation logic is tested completely. The test suite should drive a system through all logical system states to exercise all possible cases of operation. This is a difficult task.

There are approaches [2][6][7][8] that allow the generation of tests automatically from requirements with certain requirement coverage. However, this is possible only if requirements are specified in a formal language, which is rarely the case in an industrial setting.

When requirement coverage is not available, code coverage tools support test suite designers by providing the information about which parts of the source code are covered during system execution. When some of the code is not covered, a test engineer can extend the test suite with additional test cases to achieve the desired level of coverage. The goal is to reach a level of certainty that the tests did not leave out some code that contradicts the requirements.

Unfortunately, code coverage tools only produce source code coverage information. For a test engineer it is often hard to understand what the noncovered parts of the source code do and how they relate to requirements. The situation is even more complex when code generation tools are used to generate a large part of the source code as is common in industry. In these cases even programmers may find it difficult to interpret the code coverage information. Even when source code is hand-written, mapping it to requirements is not trivial. This is because requirements are typically concerned with system operation rather than with source code implementation. Many transformations applied to source code before it becomes an executing system complicate relating the requirements to source code.

To solve this problem, we have developed an approach to provide design coverage information. We instrument the code to obtain traces of system operation and then we relate these traces to design diagrams. A trace of the instrumented code execution is analyzed and the results are visualized as coverage of design diagrams. We believe that our approach simplifies the construction of test suites that achieve a desired level of design coverage.

A design coverage tool makes it unnecessary for test engineers to analyze the source code in order to understand which parts of the design are not covered by a test suite — a task that may require deep understanding of the source code and its relation to the design. Our approach is also effective when source is generated by code generators or is otherwise difficult to understand or to statically map to the design, as is in the cases of event-based designs, table-driven implementation of state machines, or other models with extensive run-time binding of components. At the same time design coverage tools do not substitute for code coverage tools when detailed coverage data of certain code is needed.

This paper is an initial concept paper, so it does not provide evaluation of the design coverage usefulness or effectiveness of its implementation. This is an area for future research.

The next section describes the idea of design coverage in more detail. In sections 3-6 we present the coverage of common design abstractions. We then describe our prototype implementation of design coverage tools. We finish with a discussion of related work and conclusions.

## 2 DESIGN COVERAGE USING THIRD EYE

Figure 1 illustrates the elements of a typical software development process that are relevant for our discussion. The software is designed to satisfy the requirements whether or not requirements exist in a tangible form before the system design is completed. Source code is produced by programmers interpreting the design documents and by code generation tools that generate code from higher-level design descriptions. The code is compiled and otherwise transformed to produce an executable. A test suite is designed to validate the implementation of the system against the requirements. It provides an environment for controlled execution of the system in which the validation can be performed. A code coverage analyzer monitors the instrumented system in operation to determine which elements of the source code were covered during execution.

Current code coverage tools force test engineers to understand well the structure and elements of the implementation. As discussed in [19], the structure and elements of the source code may be very different from the structure and elements of the executing system. Most of the requirements, as a rule, are concerned with the system operation. It could be sufficient for test engineers to concentrate only on the execution structure. However today in order to develop a complete test suite that covers all different abstract states of system operation test engineers must rely on code coverage tools, which mainly relate the coverage to source code. Thus test engineers are forced to learn more about details of software implementation than is really necessary.

Design partly describes system operation and partly system implementation in the source code. In this way the design both maps onto the executing system and provides a high-level mapping of the execution structure to the source code. Test engineers can learn much about the system execution from its design documents. If it were possible to determine the design coverage from the system execution, test engineers could use more design level information in building complete test suites.

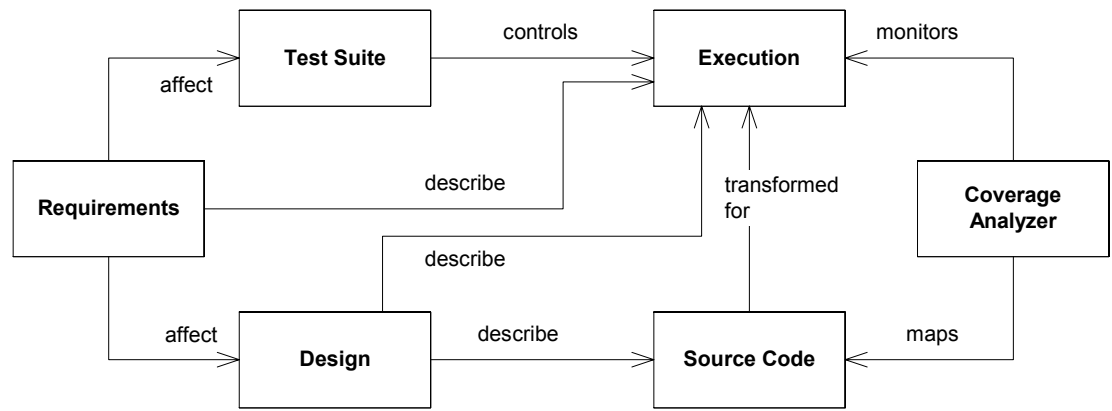

**Figure 1. Code coverage in software development process**

The design coverage can be obtained using execution tracing tools with code instrumented to report events corresponding to abstractions in system design documents. A trace tool analyzes the semantics of a design to identify events corresponding to design elements (Figure 2). Code written or generated from a design is augmented with trace events. An execution of a program produces a trace that is analyzed and mapped to the design by a trace analysis tool. Such mapping determines which elements of the design were covered and which were not covered by program execution. This is design coverage.

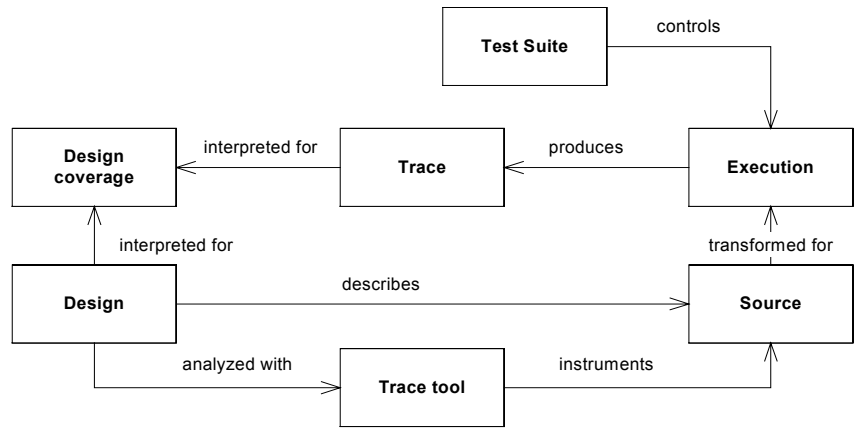

**Figure 2. Design coverage using trace tools**

Lencevicius, Ran and Yairi have developed an approach for specification-based execution tracing and analysis called Third Eye [15]. It is a methodology for tracing software execution by reporting events. "Event" is a qualitative change in the state of an entity either meaningful in the application domain or significant architecturally. When used for design coverage, events represent information that can be easily mapped to design abstractions.

In the Third Eye framework, events are typed objects. An event type has a name, a list of named and typed properties, and the type constructor. Third Eye event types are similar to classes in programming languages although the only method associated with the event type is its constructor. Events in Third Eye are characterized by the time and location of their occurrence. Event types can be defined for entering or exiting states, creating objects, initiating or completing operations, and other design level abstractions. Third Eye specifications define constraints on the properties of the events, their sequence, location, and timing. When used for design coverage, the constraints map event traces to the semantic description of a design.

We use Third Eye to report trace events and to analyze traces for design coverage. The implementation of design coverage is described in section 7. It is especially effective when code generation tools are used and Third Eye instrumentation is generated along with the operation code.

The implementation of design coverage in testing depends on tools and descriptions used for design. The sections 3-6 present design coverage for common design models.

## 3 STATE MACHINE BASED DESIGNS

One way to model system behavior is using a state machine model [3]. In this model, the abstractions are states, events, and transitions. Transitions are caused by internal or external events such as values of internal structures, user input or real-time events.

When analyzing a state-machine based design, test engineers need to know which states, events, and transitions occurred in the test. Consequently, the design coverage of state machines consists of state, event, and transition coverage. A test covers a state if it enters the state; it covers an event if the event associated with a state occurs in the state, and it covers a transition if this transition is taken during the test. This is the coverage criterion for statecharts. There are other possible criteria, however here and for other design models we present the criteria that we believe most closely relate to the model semantics and therefore are most appropriate. Test engineers may choose other criteria if necessary.

We demonstrate the coverage for a state-machine model with a simplified mobile phone application statechart (Figure 3). The application starts in the initial Idle state from which the application can transition to the Editing state where the user is editing an SMS or to the Dialing state, where the user is dialing a call. The SMS editing can be interrupted by an incoming call, at which telephone transfers to the Talking state. Finally, after hanging up, the telephone returns to the Idle state.

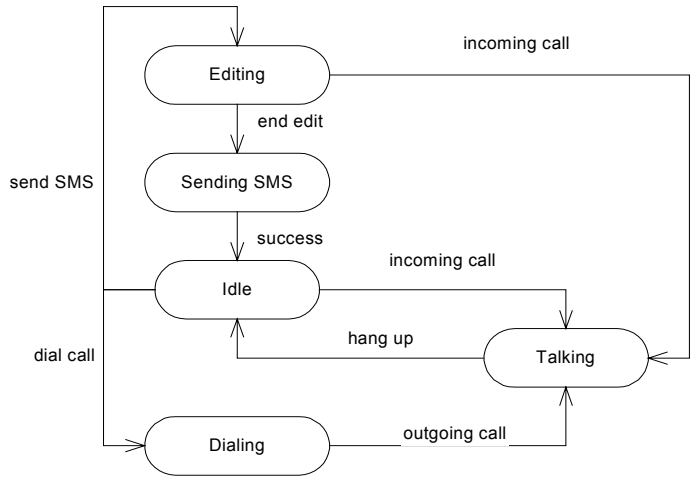

**Figure 3. Mobile phone application statechart**

This diagram expresses important requirements of precedence of the incoming call over editing an SMS message and the need to complete SMS send before an outgoing call can be dialed. Although the diagram is very simple, the code that implements the corresponding control structure may be very complex because state transitions depend on external events and are integrated with different subsystems of the mobile phone. Consequently, the information obtained from traditional code coverage tools would be very difficult to use and would require the test engineer to deeply understand the structure of the event-based phone software. It is hard to make decisions about coverage of the original design from the code coverage in the test. Using our approach, the test engineer receives coverage information that corresponds directly to the high-level diagram above. For example, assume the test scenario in which the user edits an SMS message and sends it. Since design coverage provides state, event, and transition coverage, it shows the visited states, events that occurred, and transitions taken in the test (Figure 4). Covered states are grayed out, events and covered transitions are bolded providing coverage visualization. Similar diagram would be shown in a design coverage tool integrated in a design environment. Tools could show coverage incrementally during program execution or summarized after the execution.

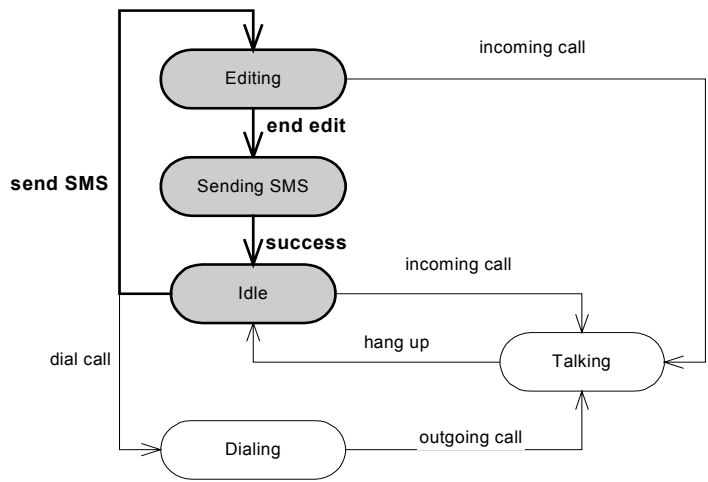

**Figure 4. Mobile phone application statechart design coverage**

An engineer can use design coverage information to select tests for states and transitions not covered in the current test suite. In our example, the coverage shows that Dialing and Talking states as well as some transitions are not covered, so the engineer can design a test that uses dial call, outgoing call and other events to cover these states and transitions.

To obtain design coverage, events occurring during execution time have to be matched with the semantic representation of the design model. We propose a design representation and mapping process for the statecharts that is simple to implement as a proof of concept, but is not optimized for performance. A statechart is represented as a tuple database, where tuples have the form transition(InitialState, FinalState, Event). For example, the statechart above is represented as:

>  transition(Idle, Talking, incoming call);
>  transition(Idle, Editing, send SMS);
>  transition(Idle, Dialing, dial call);
>  transition(Editing, Sending SMS, end edit);
>  transition(Editing, Talking, incoming call);
>  transition(Sending SMS, Idle, success);
>  transition(Dialing, Talking, outgoing call);
>  transition(Talking, Idle, hang up);

To obtain design coverage, a program would be instrumented with state(CurrentState) events at the points where the program is in the CurrentState and with event(CurrentEvent) events at the points where CurrentEvent occurs. An execution of the SMS send example would produce the following trace:

>  state(Idle);
>  event(send SMS);
>  state(Editing);
>  event(end edit);
>  state(Sending SMS);
>  event(success);
>  state(Idle);

This trace is matched with the statechart representation above. The matching can be done using a state machine traversal. The matching algorithm finds a tuple in the statechart representation that corresponds to the current state and the event following it in the trace. The algorithm determines the next state from the found tuple without enforcing execution correctness—if the next state determined from a state-machine traversal differs from the next state found in the trace, the tool stops and informs the user that a mismatch occurred. The result of the matching algorithm is all design tuples matched against the trace.

Information from these tuples is used to highlight states and transitions in the graphical statechart representation (Figure 4).

## 4 ACTIVITY DIAGRAMS

Activity diagrams are another way to model system behavior [3]. Activity diagrams are composed from states (including initial, action and final states), branches, concurrent forks and joins, transitions, and object flow indicators.

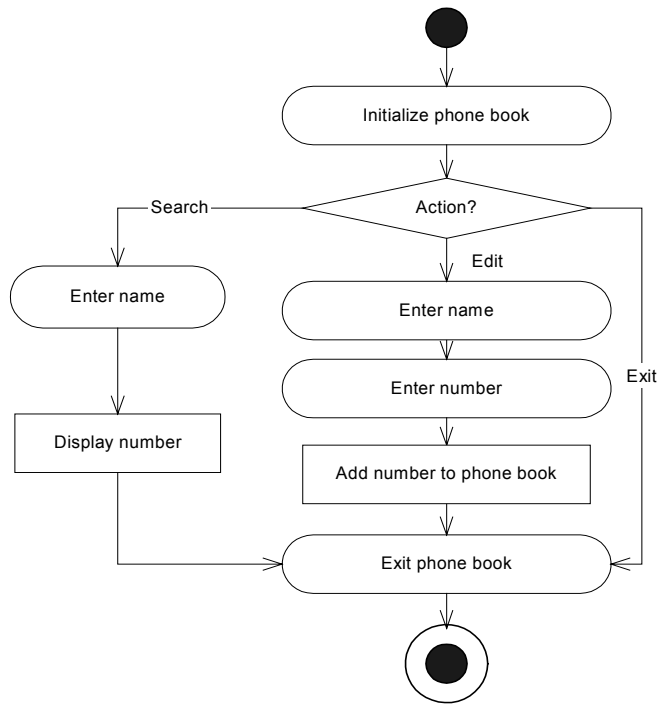

**Figure 5. Phone book activity diagram**

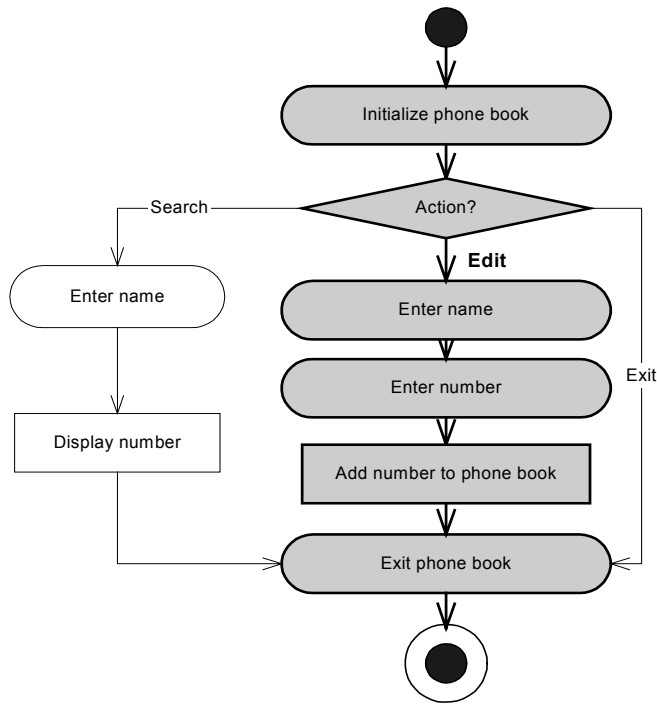

**Figure 6. Phone book activity diagram coverage**

Design coverage of activity diagrams represents the high-level control flow of an application. The only item of an activity diagram that influences coverage is a branch. The execution of a program and the elements of activity diagram covered depend on which outgoing transition from a branch is taken. All blocks between two branches form a sequence that is executed in full or not at all—it is similar to a basic block at programming language level. Though there may be a lot of code not covered during execution inside these blocks, the design coverage is decided by the branch direction. To make sure that all blocks between two branches are executed in a system with interrupts and exceptions, design coverage tools check that both an "entry" branch and the "exit" branch of a sequence are present in the trace. A test engineer can obtain a test suite covering the activity diagram by selecting use cases that follow different outgoing branch transitions. Design coverage using branches in activity diagrams is similar but not equivalent to branch coverage in source code, since it is done at the design level.

Activity diagrams are represented using branches and their outgoing transitions as follows: decision(BranchID, TransitionID1:nextBranch1, TransitionID2:nextBranch2, …). For example, the activity diagram in Figure 5 that has only one branch is represented as decision(Action?, Search:End, Edit:End, Exit:End). The code would be instrumented to report branches as branch(BranchID) and transitions taken as transition(TransitionID). The matching would be performed similarly to statechart matching, branch events corresponding to states and transition events to transitions.

Consider the activity diagram for a simplified phone-book application (Figure 5). It allows user to select a search in the phone book, to add a new number, or to exit. The coverage of a test case that adds a number to the phone book is shown in Figure 6. The states and branches are highlighted, while transitions are shown with heavy arrows. The complete test suite would also need to exercise two other branch transitions: the search capability and the straight exit, since they are not covered in this test case.

## 5 MESSAGE SEQUENCE CHARTS

Yet another way to model system behavior is by using message sequence charts [3]. Message sequence charts are composed of the actors (or objects) placed horizontally and messages between actors placed according to their time order vertically.

A message sequence chart is covered if during the program execution all messages in the chart are sent between the same actors and in the same order as they occur in the chart. This coverage criterion requires strict ordering of messages. The order of message transmissions is determined from the message timestamps. The message sequence charts can be represented as a set of tuples sends(actorSender, actorReceiver, message, eventID) and follows(eventFirstID, eventNextID).

The following fragment of the WAP (Wireless Application Protocol) [21] client implementation illustrates design coverage of message sequence charts.

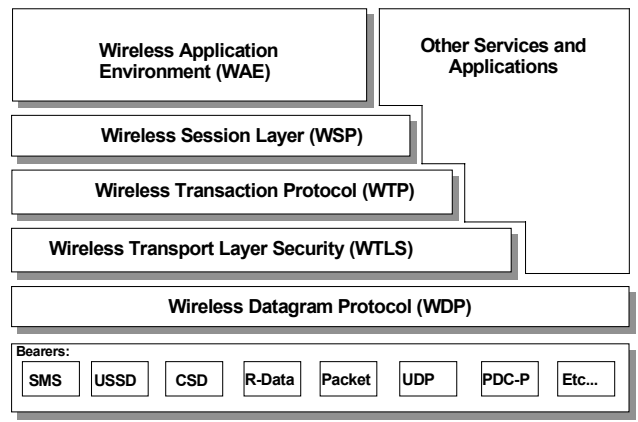

**Figure 7. WAP layered architecture**

The Wireless Application Protocol (WAP) is an industrial standard for applications and services that operate over wireless communication networks. WAP specifies an application framework and network protocols for wireless devices such as mobile telephones, pagers, and personal digital assistants (PDAs). WAP layered architecture (Figure 7) provides an application layer with a Wireless Session Protocol (WSP) interface to session services. A connection-oriented service operates above the Wireless Transaction Protocol (WTP) layer.

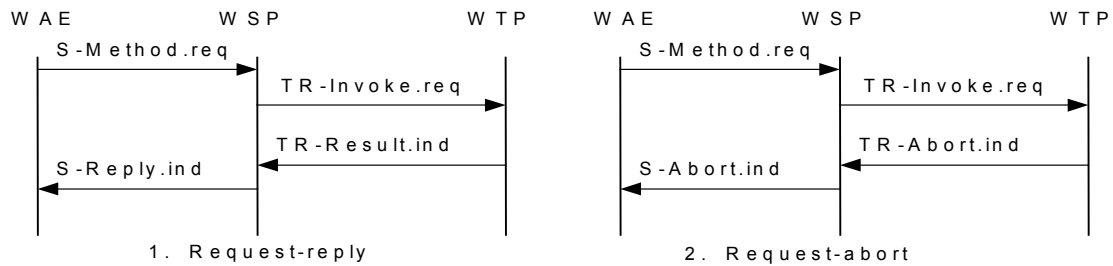

**Figure 8. WAE-WSP-WTP sequences for request-response**

When application requests information through the WAP protocol, such request passes through protocol layers as a sequence of primitives defined by the WAP standard. The primitives defined in the standard correspond to implementation actions. Specifically, a request for information by an application (WAE) is handled as a method invocation at the session layer (WSP). In turn, this method invocation is translated into a transaction at the transaction layer (WTP) (Figure 8). In detail, the invocation of an S-

Method request is followed by the invocation of the TR-Invoke request. When the transaction returns a TR-Result indication with a result, it is used in S-Reply that forwards received information to the application. This message sequence chart is shown on the left-hand side of the Figure 8 (chart 1). The right-hand side shows a message sequence chart in which the information request was aborted (chart 2). We have simplified both message sequence charts for clarity purposes. Full message sequences contain additional confirmations and acknowledgements.

The representation of the chart 1 is:

```
sends(WAE, WSP, S-Method.req, e1);
sends(WSP, WTP, TR-Invoke.req, e2);
sends(WTP, WSP, TR-Result.ind, e3);
sends(WSP, WAE, S-Reply.ind, e4);
follows(e1, e2);
follows(e2. e3);
follows(e3, e4);
follows(e4, e5);
```

The test may follow either of message sequence charts in Figure 8. If the test followed the message sequence chart 1 (shown on the left-hand side of Figure 8), the trace would be:

```
send(WAE, WSP, S-Method.req);
send(WSP, WTP, TR-Invoke.req);
send(WTP, WSP, TR-Result.ind);
send(WSP, WAE, S-Reply.ind);
```

This trace is matched against the design description by matching the actors and a message in a send event against sends tuples. The algorithm determines which sends tuples could follow the matched sends tuple and marks these as possible *next* tuples. The next event in a trace is compared against the possible *next* tuples and the process is repeated. If during execution the algorithm matches all sends tuples in a single MSC representation, this MSC is covered by the execution. In the current example, the trace matches with the design representation of the left message sequence chart, so this chart is covered. Then the test engineer could device a test that covers the MSC on the right by providing abort condition from the WAP server.

## 6 STRUCTURE DIAGRAMS

Structure diagrams model the structure of a system and not its behavior. Structure diagrams include class, object, component and deployment diagrams. Design coverage in structure diagrams focuses on the high-level runtime configuration of objects. Concrete object configurations are shown in object diagrams, but possible associations between objects in the current practice of UML are shown in class diagrams. Class diagrams contain a set of classes, interfaces and their relationships. Consider a simple class diagram in Figure 9 that shows a collaboration in call forwarding.

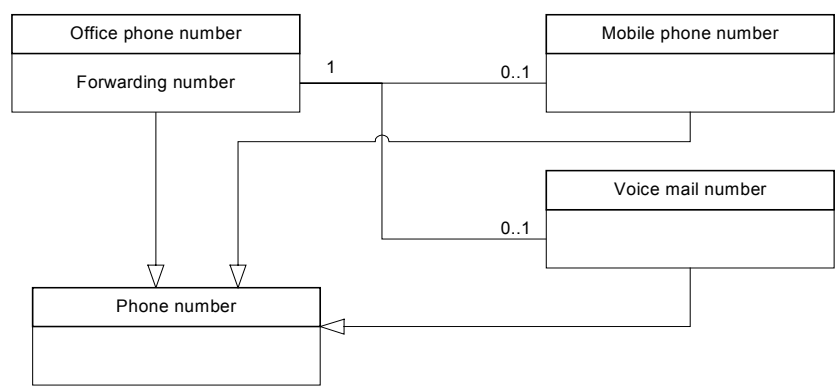

**Figure 9. Call forwarding class diagram**

In this diagram, an Office phone number class has a field Forwarding number. The forwarding number can be associated with zero or one mobile phone numbers or voice mail numbers. All these phone number classes are subclasses of a Phone number class.

What is the criterion of coverage for a class diagram? Generalization constraints between classes are static and not subject to test coverage. However, associations are instantiated dynamically during program execution and yield different object *configurations*. Design coverage indicates what *configurations* of associations were covered during a test. For example, if a test instantiated one mobile phone number and no voicemail numbers, then the test only covers this configuration, but not the configuration with one mobile phone number and one voicemail number. Design coverage information allows a test engineer to design additional tests that construct different object configurations.

The association coverage implementation requires detecting object configurations that occur during run time. In our example, it is necessary to check whether the field containing forwarding numbers was assigned mobile phone number and voicemail phone number. In more complex relationships objects may create a complicated graph of references. Finding occurrences of such graphs at run time is an NP-complete problem [12], because the example graph has to be matched against the whole program object graph. In addition such matching has to be done after each field assignment that can change the object graph. One of the authors has previously implemented a tool called query-based debugger [12][13] that has the functionality needed and is optimized for the task. This tool can be used to detect structure diagram coverage.

## 7   IMPLEMENTATION

Design coverage is best implemented as a service in an integrated design and programming environment that has tracing capabilities. The implementation needs to trace design elements during the test execution and map the reported events back to the design. These requirements are best accomplished in an environment that already closely ties the design and the code. In fact, environments such as I-Logix Rhapsody [9] and Rational Rose RealTime [20] have some tools that resemble design coverage. However, the design coverage can be implemented in practically any design and programming environment by using tracing tools. In the worst case, programmers need to manually instrument the code to insert trace events. Such manual instrumentation may provide incorrect or inconsistent information, diminishing the usefulness of the design coverage system. Yet sometimes this instrumentation can be piggybacked onto tracing events inserted by programmers for other purposes, reducing the effort and providing more reliable information.

We have implemented a prototype design coverage tool for a proprietary CASE environment that is used at Nokia to design and generate the source code for mobile telephone user interface software. This CASE tool uses state-machine paradigm with event and real-time based state transitions. Third Eye tracing and trace analysis framework [15] provided us with facilities to report structured events needed for coverage information tracing.

We instrumented the code generated by our proprietary CASE tool to report traces[1]. The UI application code contains a function that handles state transitions. Third Eye trace function calls are inserted into the switch statement of this function for each case that corresponds to the state transition in the design diagram. Trace calls set events' properties: the application ID, the current and new state names, the transition ID, and the timestamp. Property values are obtained from the tool internal information. The events are reported during application's execution and supply all information needed to map the transition back to the design diagram. An analysis tool, that we have not implemented yet, would visualize design coverage as given in examples. Examples of design coverage for mobile phone applications are given in [14].

Our implementation of design coverage for the message sequence charts is also based on the Third Eye framework. However, since WAP code was hand written and not generated, we have added trace function calls in the places that correspond to message transmissions. The trace events are collected, analyzed and mapped back to the message sequence charts as described in section 5.

We have not yet implemented the design coverage for other design abstractions.

In our experience, the instrumentation did not noticeably slow down executing applications. The overhead of the design coverage instrumentation should be lower than the code coverage overhead, since traced events are less frequent. The matching overhead depends on the abstractions used. It is polynomial in terms of the number of tuples describing designs and linear in terms of the number of tuples in the trace for statecharts, activity diagrams and message sequence charts. A detailed study of the overhead should be done in the future.

## 8    RELATED WORK

Some integrated design and programming environments such as I-Logix Rhapsody [9] and Rational Rose RealTime [20] have tools that resemble design coverage. For example, in Rhapsody and Rose RealTime, test engineers can follow the test's execution on the design diagrams. However, no tools contain a dedicated design coverage tool and no tools contain structure diagram coverage. Furthermore, we propose general design coverage methods instead of ad hoc implementations for some design abstractions, for example, just state transition diagrams. Consequently our approach can be applied in a generic way for all design abstractions and implemented for a wide variety of environments.

If requirements are specified in a formal language, there are approaches [2][6][7][8] that generate tests automatically from requirements with certain requirement coverage. Unfortunately, formal requirements are rarely available in industrial setting. Design coverage can be obtained for all tests even for the ones created manually. "Specification-based coverage analysis" mentioned in the papers [2][6][7] means the analysis of a specification to determine whether all its elements have a test vector. This is orthogonal to our approach. Design coverage enables analysis of a test to determine which elements of design do not have a test vector.

The mapping between design and source code and between design and execution has been extensively researched from different angles. E.W. Dijkstra [5] and M. Jackson [11] among others have advocated the mapping between a problem and the program source code. Design coverage uses mapping from program execution traces to the design model. Reverse engineering research [17] aims to produce design models from source code or from execution traces. G.C. Murphy, D. Notkin and K. Sullivan [18] propose matching of an extracted source model with the user supplied model to find similarities and differences. J. Andrews and Y. Zhang propose a way to validate software by matching the log files against the system's parameterized state machine specification [1]. A similar method can be used to match traces against the statecharts.

---

[1] A simple modification of the CASE tool's code generator would insert state transition events into the generated code automatically.

Structure diagram coverage is similar to problems addressed by shape analysis in programming language field [4]. However, shape analysis only produces a set of possible structure shapes and summarizes some information. It does not guarantee coverage of a certain shape during execution. Jackson and Waingold developed a tool called Womble [10] that statically extracts object models from Java code. This tool also produces only possible object relationships and does not guarantee shape occurrence during execution.

C. Liu and D. Richardson [16] introduce a concept of application states as a set of property/value pairs interesting to the test engineer during a particular test. They propose application state coverage metric as a ratio of covered application states to total number of application states. They do not tie the application state to the design, so their idea is complementary to the design coverage.

## 9 CONCLUSIONS

This paper deals with a problem of relating the test data to the design in a way similar to code coverage. Availability of such design coverage would simplify the design of a test suite that covers all necessary project requirements. Code coverage fails to achieve similar result in two ways: it may indicate that non-essential code was not covered or it may leave the testing engineer wondering whether an important design element was exercised.

We have proposed a new generic approach and a prototype tool implementation that directly produces design coverage of the executed test suite. This approach simplifies the design of test suites by not requiring a test engineer to understand low-level details of the source code. It also satisfies the need to check the coverage of design abstractions. Using design coverage, test suites can be created faster by focussing on untested design elements.

We have demonstrated design coverage for common design abstractions such as state machines, activity diagrams, message sequence charts and structure diagrams. Our prototype tool implementation allowed us to obtain the design coverage for mobile phone applications generated by a proprietary CASE tool. This is an initial concept paper, so additional research is needed on effectiveness and efficiency of the design coverage in practice.

## 10 ACKNOWLEDGEMENTS

We thank Andre Dolenc and Andy Turner for support of this research. We thank Gopal Raghavan for valuable comments on this paper.